\documentclass[onecolumn,journal]{IEEEtran}
\usepackage{cite}
\usepackage{amsmath,amssymb,amsfonts}
\usepackage{algorithmic}
\usepackage{graphicx}
\usepackage{textcomp}
\usepackage{subcaption}
\usepackage{caption}
\usepackage{balance}
\usepackage{wrapfig}
\usepackage{algorithm}
\usepackage{xcolor}
\usepackage{float}
\usepackage{physics}
\usepackage{setspace}
\usepackage{comment}

\def\BibTeX{{\rm B\kern-.05em{\sc i\kern-.025em b}\kern-.08em
    T\kern-.1667em\lower.7ex\hbox{E}\kern-.125emX}}
\begin{document}
\title{Feature-Based Generalized Gaussian
Distribution Method for NLoS Detection in Ultra-Wideband (UWB) Indoor Positioning System}
\author{Fuhu Che, Qasim Zeeshan Ahmed, Jaron Fontaine, Ben~Van~Herbruggen, Adnan~Shahid, Eli~De~Poorter, and Pavlos I. Lazaridis
\thanks{Part of the paper is also presented in 26th IEEE International Conference on Automation and Computing (ICAC), 2021. This work is supported in part by RECOMBINE-MSCA-Research and Innovation Staff Exchange (RISE)-2019 under Grant ID: 872857, the Capacity Building for Digital Health Monitoring and Care Systems in Asia EAC-A02-2019-CBHE under Grant ID: 619193, the Fund for Scientific Research Flanders, Belgium, FWO-Vlaanderen, FWO-SB, under Grant 1SB7619N, and the imec AAA UWB localisation research project 75.}
\thanks{Fuhu Che, Qasim Zeeshan Ahmed and Pavlos I. Lazaridis are with School of Computing and Engineering, University of Huddersfield, Huddersfield, UK (e-mail: q.ahmed@hud.ac.uk). }
\thanks{Jaron~Fontaine, Ben~Van~Herbruggen, Adnan~Shahid, Eli~De~Poorter are with IDLab, Department of Information Technology at Ghent University - imec, Belgium.}
}
\doublespacing
\IEEEtitleabstractindextext{%
\fcolorbox{abstractbg}{abstractbg}{%
\begin{minipage}{\textwidth}%
\begin{abstract}
Non-Line-of-Sight (NLoS) propagation condition is a crucial factor affecting the precision of the localization in the Ultra-Wideband (UWB) Indoor Positioning System (IPS). Numerous supervised Machine Learning (ML) approaches have been applied for NLoS identification to improve the accuracy of the IPS. However, it is difficult for existing ML approaches to maintain a high classification accuracy when the database contains a small number of NLoS signals and a large number of Line-of-Sight (LoS) signals. The inaccurate localization of the target node caused by this small number of NLoS signals can still be problematic. To solve this issue, we propose feature-based Gaussian Distribution (GD) and Generalized Gaussian Distribution (GGD) NLoS detection algorithms. By employing our detection algorithm for the imbalanced dataset, a classification accuracy of $96.7\%$ and $98.0\%$ can be achieved. We also compared the proposed algorithm with the existing cutting-edge such as Support-Vector-Machine (SVM), Decision Tree (DT), Naive Bayes (NB), and Neural Network (NN), which can achieve an accuracy of $92.6\%$, $92.8\%$, $93.2\%$, and $95.5\%$, respectively. The results demonstrate that the GGD algorithm can achieve high classification accuracy with the imbalanced dataset. Finally, the proposed algorithm can also achieve a higher classification accuracy for different ratios of LoS and NLoS signals which proves the robustness and effectiveness of the proposed method.
\end{abstract}

\end{minipage}}}

\maketitle

\section{Introduction}
\label{sec:introduction}
\IEEEPARstart{W}~ith the rapid development of the Internet of Things (IoT)s, the requirement of a precise indoor positioning system (IPS) has attracted considerable attention in the research community and industry~\cite{ Pinto-2021,Jiang-2021, Ahmed-2021a, Wu-2021, Fred-2022, YWu-2021, Alsmadi-2021, Ahmed-2020,Abou-2021,Ahmed-2013}. Several examples aforesaid, pedestrian tracking systems~\cite{YWu-2021, Pu-2021, Wang-2021}, autonomous flying drones in warehouses~\cite{Hayat-2016, Chhikara-2021,Wang-2021}, and social distancing requirements caused by pandemic such as COVID-19~\cite{Nguyen-2020, Chamola-2021}, etc., require accurate IPS. The Global Navigation Satellite System (GNSS) provides tremendous convenience to human life as they provide real-time localization in open space. Unfortunately, the GNSS signals are attenuated severely by the wall and fail to achieve the accurate positioning in indoor environments~\cite{Jiang-2021,Kbayer-2018}. Among various indoor positioning technologies, Ultra-wideband (UWB) can achieve high accuracy due to its characteristics of extremely short pulse that provides good time resolution ~\cite{Wang-2021,Hanssens-2018,Ahmed-2020,Ahmed-2007}. However, the accuracy of UWB IPS could be significantly affected when the NLoS signal occurs~\cite{Wang-2021a, Che-2020, Waqas-2021}. The NLoS condition exists when the signals between the transceivers are reflected or blocked by the obstacles. In this case, signal propagation delay occurs, resulting in longer Time-of-Flight (ToF) and an estimated distance error between the transmitter and receiver~\cite{Che-2020,Waqas-2021}. Thus, significantly reducing the accuracy of IPS. 

The current literature includes several research works that enhance the accuracy of the UWB IPS by identifying whether the signal has a LoS or NLoS component~\cite{Wang-2021,Hanssens-2018,Che-2020,Yan-2017}. The methods of NLoS identification can be coarsely summarised into two types (i) Non-feature based NLoS signal classification that uses context information and (ii) feature-based NLoS identification relying on the UWB waveforms. The \textit{non-feature based approach} from~\cite{Yan-2017} uses a modified Kalman filter to classify LoS/NLoS conditions based on the Bayesian sequential of range measurements. In~\cite{Gururaj-2017}, the authors present a real-time NLoS identification approach based on the received signal strength without the training phase and prior knowledge of the environment. In~\cite{Abolfathi Momtaz-2018} models the NLoS as a deterministic additive term and identifies NLoS based on the statistical features of range measurements. In~\cite{Tian-2020}, a fusion technique such as an Inertial Navigation System (INS) is combined with UWB for pedestrian tracking. In~\cite{Wang-2021}, the authors apply the floor map injunction with INS and UWB to predict the state and then determine and recognise the NLoS signals. 

In contrast, in the \textit{feature-based} NLoS \textit{identification}, the UWB waveform signals under the LoS are different from those under NLoS conditions. These features can be extracted from the UWB signal to identify NLoS conditions by employing Machine Learning (ML) algorithms. One of the early ML-based NLoS identification approaches in UWB was proposing the Support Vector Machine (SVM) algorithm as a classifier in~\cite{Marano-2010,Wymeersch-2012}. In these papers~\cite{Marano-2010,Wymeersch-2012,Chitambira-2017}, the identification of LoS and NLoS signals was considered as a binary classification problem. The results proved that the ML approaches could improve the accuracy of UWB IPS by identifying the NLoS signals. Different ML techniques like Naive Bayes (NB)~\cite{Che2-2020}, Boosted Decision Tree (BDT)~\cite{Krishnan-2018}, etc., were also investigated. Deep-learning based approach such as Convolutional Neural Network (CNN) was developed in~\cite{Zheng-2020}. Furthermore  in~\cite{Fontaine-2020}, the authors propose a semi-supervised based ML approach using autoencoders which achieves $29\%$ higher accuracy than state-of-the-art deep neural network algorithm. However, the above-mentioned feature-based methods have drawbacks, especially when the data is imbalanced and a small number of NLoS data samples are present. In such cases, it is hard for such algorithms to train a robust classifier for NLoS identification. To address this shortcoming, we propose a Gaussian Distribution (GD) and Generalized Gaussian Distribution (GGD) algorithms for NLoS signal detection in the presence of imbalance datasets. Our proposed GGD is an unsupervised learning algorithm. We start with the LoS data to build a training dataset, determine a threshold through density estimation according to the GGD of each feature, and test the classification of the new data according to this calculated threshold. 

Therefore, the main contributions of this paper are as follows.
\begin{itemize}
    \item Study the performance of an unsupervised learning algorithm based on Gaussian Distribution (GD) and Generalized Gaussian Distribution (GGD) algorithms to discriminate between LoS and NLoS conditions in presence of imbalance datasets with limited NLoS training data for IPS.
    \item Compare the proposed algorithm with the existing supervised ML algorithm (SVM, DT, NB, and NN) in terms of the confusion matrix, receiver operating characteristics (ROC) curve and the area under the curve (AUC) to show the superior performance of the proposed algorithms.
\end{itemize}

The remainder of this paper is organized as follows. Section~\ref{sec:Section-2} describes the overall UWB system model and the principle of UWB localization. In Section~\ref{sec:Section-3}, our proposed classification algorithms are discussed. In Section~\ref{sec:Section-4}, the principle of our proposed
algorithms is presented followed by the features used for NLoS
signal classification. In addition, this section also discusses the environment in which the data was collected, and the hardware used for this data collection. Section~\ref{sec:Section-5} presents the performance evaluation of the proposed algorithms and compare the results with the state-of-art ML algorithms in detail. The summary of the accomplishment is given in Section~\ref{sec:Section-6}.

\section {ABBREVIATIONS AND ACRONYMS}
\begin{tabular}{l l}
Angle-of-Arrival &- AOA\\
Area under the curve &- AUC \\
Additive White Gaussian Noise &- AWGN \\
Boosted Decision Tree &- BDT\\
Channel Impulse Response &- CIR\\
Conventional Neural Network &- CNN\\
Comma-Separated Value &- CSV\\
False Negative &- FN\\
False Positive &- FP\\
False Positive Rate &- FPR\\
Gaussian Distribution &- GD  \\
Generalized Gaussian Distribution &- GGD \\
Global Navigation Satellite System &- GNSS \\
Inertial Navigation System &- INS\\
Internet of Things &- IoTs\\
Indoor Positioning System &- IPS\\
Line-of-Sight &- LoS\\
Import Vector Machine &- IVM \\
K-Nearest Neighbor &- KNN \\
Multi-Layer Perceptron &- MLP\\
Naïve Bayes &- NB	\\
Non-Line-of-Sight &- NLoS	\\
Probability Distribution Function &- PDF  \\
Receiver Operating Characteristics  &- ROC \\
Support Vector Machine &- SVM\\
Time-of-Arrival &- ToA\\
Time-of-Flight &- ToF \\
True Positive &- TP   \\
True Positive Rate &- TPR\\
Ultra-wideband &- UWB 
\end{tabular}

\section{UWB Positioning System Model}~\label{sec:Section-2}
\subsection{Transmitted UWB Signal by the Anchor}
We consider an UWB signal waveform $s(t)$ transmitted by the help of $K$ pulses $p$ with a period of $T_{p}$ that consists of transmitted frames~\cite{Ahmed-2015}. As the transmitted UWB signal location is known the transmitted signal is modelled as~\cite{Ahmed-2008,Ahmed-2008a}
\begin{equation}\label{eq-1}
s(t)=\sqrt{E_s}\sum_{k=1}^{K-1}{p(t- kT_p)},
\end{equation}
\noindent
where $E_s$ is the energy of the UWB signal $s(t)$. 
\subsection{Received UWB Signal by the Mobile Node}
The transmitted signal $s(t)$ experiences multipath channel effects and the received signal at the $i$-th mobile node can be expressed as~\cite{Ahmed-2011,Ahmed-2010}.
\begin{equation}\label{eq-2}
r_i(t)=\sum_{v_i=1}^{V_i}{h_{v_i}s(t-\tau_{v_i})+n(t)}, \quad i=1,2,\cdots, N.
\end{equation}
where $V_i$ is the maximum number of multipath experienced by the $i$-th mobile node. $h_{v_i}$ and $\tau_{v_i}$ represent the amplitude and delay of the $v$-th path respectively at the $i$-th mobile node, and $n(t)$ is the Additive White Gaussian Noise (AWGN) with zero mean and two-sided power spectral density $N_0/2$. However, as this $i$-th mobile node or referred to a tag in literature will be moving, therefore, our main interest will be to calculate the distance between anchor node and the $i$-th mobile node. The distance calculation is discussed in the upcoming subsection. 

\subsection{Localization System}
UWB-based IPS consists of two different kinds of nodes. Nodes with a known position are called anchors whereas the nodes with unknown position are tags and their position is to be determined. Firstly, time of arrival (ToA) technique can be used to measure the distance between the anchor and tag. Secondly, the triangulation technique helps to determine the position of the tag in a $2$-dimensional ($2D$) environment with three or more than three anchors.
\begin{figure}[!htb]
     \centering
    \begin{subfigure}[t]{0.45\textwidth}
        \raisebox{-\height}{\includegraphics[width=\textwidth]{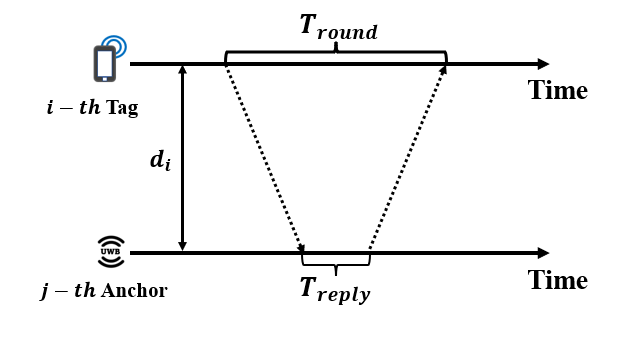}}
        \caption{Propagation time calculations}
        \label{fig-1a}
    \end{subfigure}
    \hfill
    \begin{subfigure}[t]{0.45\textwidth}
        \raisebox{-\height}{\includegraphics[width=\textwidth]{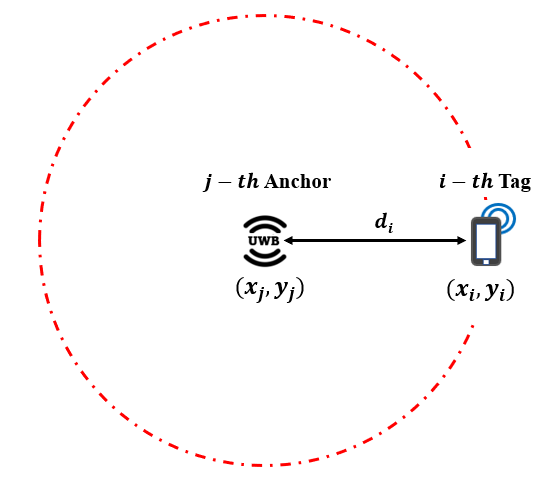}}
        \caption{Range scheme}
        \label{fig-1b}
    \end{subfigure} 
     \caption{UWB localization theory in a $2$-dimensional ($2D$) environment}
     \label{fig-1}
\end{figure}
\begin{enumerate}
\item Time of Arrival (ToA) Approach

The first process of ToA will require both the anchor and the tag to have synchronized clocks. As shown in Fig.~\ref{fig-1a}, a timestamp will be sent by the $i$-th tag to the $j$-th anchor. The timestamp will be processed by the $j$-th anchor in $T_{reply}$ seconds and send back to the $i$-th tag. The time taken for the $i$-th tag is $T_{round}$ and the propagation time $\mathrm{\tau_{i,j}}$ can be expressed as
\begin{equation}\label{eq-3}
\mathrm\tau_{i,j}= \frac{T_{round}-T_{reply}}{2}, \quad i=1,2,\cdots, N, j=1,2,3.
\end{equation}
The estimate distance $d_{i,j}$ between the $i$-th anchor and $j$-th tag can be calculated as
\begin{equation}\label{eq-4}
d_{i,j}=c \times \mathrm{\tau_{i,j}},
\end{equation}
where $c$ is the speed of light in meter per seconds ($m/s$). Fig.~\ref{fig-1b} shows the distance between the $i$-th tag and the $j$-th anchor. However, the coordinates of the tag will be unknown and for that the trilateration approach is required. 

\item Trilateration Approach for UWB Localization

Fig.~\ref{fig-1b} shows the position of the $i$-th tag with respect to the $j$-th anchor. The coordinates of the $j$-th anchor are $(x_j, y_j)$ which are already known. The coordinates of the $i$-th tag are represented as $(\hat{x}_i,\hat{y}_i)$ , where $(\hat{\cdot})$ indicates the estimate of the position. The distance between each anchor and the tag is calculated as
\begin{eqnarray}\label{eq-5}
d_{i,j}= \sqrt{(\hat{x}_i-x_j)^2+(\hat{y}_i-y_j)^2}, &~& i=1,2,\cdots,N,\nonumber\\&~&  j=1, 2, 3.
\end{eqnarray}
The position of tag $(\hat{x}_i,\hat{y}_i)$ can be determined by employing the least-squares solution.
\end{enumerate}

\section{Proposed Algorithms}~\label{sec:Section-3}
We are given a UWB training set of size $TL$ such that $\mathcal{S}=\{\pmb{s}_1,\pmb{s}_2,\cdots,\pmb{s}_{TL}\}^T$. If $t$ represents the index of the dataset $\mathcal{S}$, then $t$ will consist of $M$ features and will be represented as $\pmb{s}_t= \{s_{t,1},s_{t,2},\cdots,s_{t,M}\}$. The collection of these features are discussed in detail in the upcoming Section~\ref{sec:Section-4}. Upon collection of these features, we need to design ML algorithms such that we can classify the test data $\pmb{u}_i$ of the $i$-th tag as LoS or NLoS signals. After training the dataset on the developed positioning algorithms we could classify the output of the test data $\pmb{u}_i$ as $\{l=0~\textrm{or}~1\}$. This classification will indicate the LoS $l=0$ or NLoS $l=1$ status of the received signal. Let us now discuss the proposed positioning algorithms in detail.
\subsection{Gaussian Distribution (GD)}~\label{sub-section:GD}
Assuming the feature $s_{m}$ to be Gaussian Distribution (GD) with mean $\mu_{m}$, and variance $\sigma^{2}_{m}$ can be written as
\begin{eqnarray}\label{eq-6}
P(s_{m},\mu_{m},\sigma_{m} ^{2})\!\!\!\!&=&\!\!\!\!\frac{1}{\sqrt{2\pi}\sigma_{m}}\!\exp\left(-\frac{(s_{m}-\mu_{m})^2}{2\sigma_{m} ^2}\!\right), \nonumber\\&~&m=1,2,\cdots,M.
\end{eqnarray}
However, we will still require to calculate the mean $\mu_{m}$ and the variance $\sigma_{m}^2$ of the $m$-th feature. As the exact mean and variance of the features of the dataset is unknown, we can incorporate the provided training data $TL$ to calculate their estimates. The estimate of the mean $\hat{\mu}_{m}$ and variance $\hat{\sigma}^{2}_{m}$ for the $m$-th feature can be calculated as 
 \begin{eqnarray}~\label{eq-7,8}
 \hat{\mu}_{m} &=&\frac{1}{TL}\sum_{t=1}^{TL} s_{t,m},\quad m=1,2,\cdots,M,\\
 \hat{\sigma}_{m}^2 &=&\frac{1}{TL}\sum_{t=1}^{TL}{(s_{t,m}-\hat{\mu}_{m})}^2, m=1, 2, \cdots, M.
\end{eqnarray} 
Once we have the estimates of the mean and variance of each feature on training data, given a test data $\pmb{u}_i$ we can calculate the probability as
\begin{equation}\label{eq-9}
P(\pmb{u}_i)=\prod_{m=1}^{M}P(u_{m},\hat{\mu}_{m},\hat{\sigma}_{m}^{2}),\quad i=1,2,\cdots,N, 
\end{equation}
and classify the output as:
\begin{equation}\label{eq-10}
l = 
\begin{cases}
P(\pmb{u}_i) > \epsilon,  \quad $LoS$\\
P(\pmb{u}_i) \leq \epsilon,   \quad $NLoS$
\end{cases}
\end{equation}
where $\epsilon$ is judgment boundary and will be discussed in detail in section~\ref{sec:Section-4}.

\subsection{Generalized Gaussian Distribution, (GGD)}
By using GD algorithm, some abnormal features of LoS data may be difficult to classify, as a result the model could wrongly classify it as a NLoS component.  Furthermore, the GD algorithm require two key parameters to be modelled: a) mean and b) variance of the data as mentioned~\ref{sub-section:GD}. In such cases, the GD model may not be able to accuracately identify the NLoS dataset. Therefore, instead of GD, Generalized Gaussian Distribution (GGD) can be adopted~\cite{Ahmed-2015}. The GGD of the $m$-th feature can be written as
\begin{eqnarray}\label{eq-11}
P(s_{m}, \mu_{m}, \alpha_{m}, \beta_{m})\!\!\!\!&=&\!\!\!\!\frac{\beta_{m}}{2 \alpha_{m} \Gamma(1/\beta_{m}\!\!)}\!\!\exp\!\!\left(\!-\frac{|s_{m}-\mu_{m}|}{\alpha_{m}}\!\! \right)^{\beta_{m}}
\end{eqnarray}
where $\mu_{m}$ is the mean, $\beta_{m}$ determines the shape of the PDF, $\alpha_{m}$ is the scale parameter of the GGD
and $\Gamma(\cdot)$ is the gamma function. The variance $\sigma_{m}^2$ and the kurtosis $\kappa_{m}$ if the GGD is given as
\begin{eqnarray}\label{eq-12}
\sigma^2_{m} &=& \frac{\alpha_{m}^2 \Gamma(3/\beta_{m})}{\Gamma(1/\beta_{m})}\nonumber\\
\kappa_{m} &=& \frac{\Gamma(5/\beta_{m})\Gamma(1/\beta_{m})}{\Gamma(3/\beta_{m})^2}-3.
\end{eqnarray}
Given a dataset $\mathcal{S}$, for GGD algorithm we need to estimate the mean  $\hat{\mu}_{m}$, variance $\hat{\sigma}^2_{m}$ and kurtosis $\hat{\kappa}_{m}$ which are calculated as 
\begin{eqnarray}\label{eq-13,14,15}
\!\!\!\!\hat{\mu}_{m}\!\!\!\! &=&\!\!\!\!\frac{1}{TL}\sum_{t=1}^{TL}{s_{{t,m}}}, \quad m=1,2,\cdots,M,\\
\!\!\!\!\hat{\sigma}_{m}^2\!\!\!\!&=&\!\!\!\!\frac{1}{TL}\sum_{t=1}^{TL}{(s_{t,m}-\hat{\mu}_{{m}})}^2, \quad m=1,2,\cdots,M, \\
\!\!\!\!\hat{\kappa}_{m}\!\!\!\!&=&\!\!\!\!\frac{\frac{1}{TL}\sum_{t=1}^{TL}{(s_{t,m}-\hat{\mu}_{{m}})}^4}{\left[\frac{1}{TL}\sum_{t=1}^{TL}{(s_{{t,m}}-\hat{\mu}_{m})}^2\right]^2}\!-\!3,\!m=1,2,\!\!\cdots,M,
\end{eqnarray}
where the estimate of kurtosis $\hat{\kappa}_{m}$ can be used to measure the shape parameter $\beta_{m}$ and estimate of variance $\hat{\sigma}^2_{m}$ can help to determine the scale parameter $\alpha_{m}$ of the GGD. Now the test data $\pmb{u}$ can be employed to calculate the probability as
\begin{equation}\label{eq-16}
P(\pmb{u}_i)=\prod_{m=1}^{M}P(u_{{m}},\hat{\mu}_{{m}},\hat{\alpha}_{{m}}, \hat{\beta}_{m}),\quad i=1,2,\cdots,N.
\end{equation}
where the classification can be done with the help of (\ref{eq-10}). For the sake of clarity, the GGD algorithm is summarised in the next subsection.
\subsection{Classification Algorithm}
For the proposed classification technique, we first extract the LoS and NLoS signal features from the received dataset $\mathcal{S}$, then use distribution of each features to establish a model $P(\pmb{s}_i)$ for the $i$-th tag. After the model is built and the threshold is selected, we fit this model with the test dataset $\pmb{u}_i$ and classify whether the signal experiences a LoS or NLoS signal. 
The steps of the training and testing stage of the proposed GGD algorithm are shown in Algorithm~\ref{algorithm-1} and~\ref{algorithm-2}, respectively. As GGD is the more generalised algorithm we have only summarised it. For GD algorithm step 5 will not be required in Algorithm~\ref{algorithm-1}. For Algorithm~\ref{algorithm-2} in step-1 we will replace (\ref{eq-16}) with (\ref{eq-9}). Let us now look into the experimental setup followed by data collection, configuration of the UWB kit and key feature extractions in detail.
\begin{algorithm}[!tb]
    \caption{: Training Stage}\label{algorithm-1}
    \textbf{Input}: Collected dataset of UWB consisting of LoS and NLoS signal features.\\
    \textbf{Output}: Create model $P(\pmb{s}_i)$ for the $i$-th tag of the dataset $\mathcal{S}$.\\
    \textbf{Algorithm}
    \begin{enumerate}
        \item Initialize and pre-process the dataset. 
    \item Select part of LoS as training data.
    \item Estimate the mean $\hat{\mu}_m$,
    \item Estimate the variance $\hat{\sigma}^2_m$, and 
    \item Estimate the kurtosis $\hat{\kappa}_m$,
    \item Construct the model $P(s_m)$ for the $m$-th feature.
    \item Select the threshold $\epsilon$ by calculating $F_1$-score value.
    \item Construct the model $P(\pmb{s}_i)$.
    \end{enumerate}
\end{algorithm}

\begin{algorithm}[!tb]
    \caption{: Testing Stage}\label{algorithm-2}
    \textbf{Input}: Test dataset with a mixture of LoS and NLoS signals. \\
    \textbf{Output}: Determine whether it is LoS or NLoS and then determine the exact location of the tag or the moving node. \\
    \textbf{Algorithm:}
    \begin{enumerate}
      \item Fit the model by calculating the probability $P(\pmb{u}_i)$ of the testing data as mentioned in (\ref{eq-16}).
    \item \textbf{If} $P(\pmb{u}_i) \leq \epsilon \gets$ NLoS signal. 
    \item \textbf{Else} $P(\pmb{u}_i) > \epsilon \gets$ LoS signal. 
    \item Determine the distance between the tag and the respective anchor using (\ref{eq-5}).
    \end{enumerate}
    \end{algorithm}

\section{Experiment Setup}~\label{sec:Section-4}
\subsection{Data Extraction and Key Feature Selection Process}
%
\begin{figure*}[!ht]
\begin{minipage}[!h]{0.5\linewidth}
\begin{center}
\title*{a-Scenario 1}
\includegraphics[width=1\linewidth]{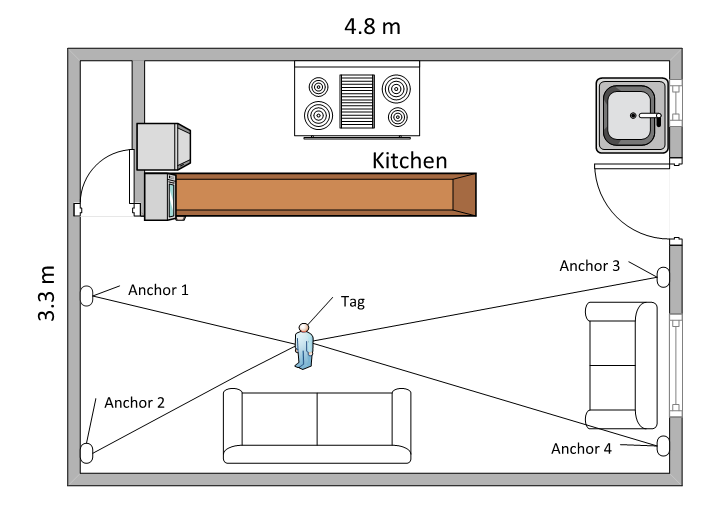}
\label{fig-2a}
\end{center}
\end{minipage}
\begin{minipage}[!ht]{0.5\linewidth}
\begin{center}
\title*{b-Scenario 2}
\includegraphics[width=1\linewidth]{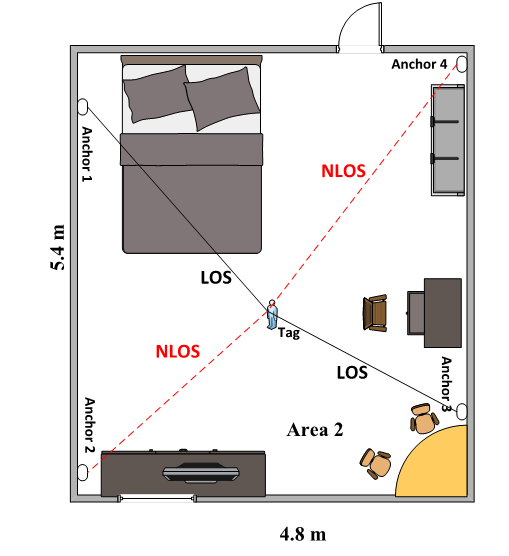}
\label{fig-2b}
\end{center}
\end{minipage}
\caption{The scenarios where training and test data were collected for evaluation:(a) Studio (b) Room}
     \label{fig-2}
\end{figure*}
In this paper, MDEK-$1001$ UWB kits from DECAWAVE company are employed for data preparation. The configuration of the UWB kit is shown in Table~\ref{table-1}. All the experiment is carried out using MATLAB (R2020b). Two separately independent datasets for evaluation were collected in a small studio environment of dimensions $(3.3 \times 4.8)$~m as shown in Figure~\ref{fig-2}(a) and a room environment of size $(4.8 \times 5.4)$~m as shown in Figure~\ref{fig-2}(b).  In this paper, we have utilised scenario-1, therefore, the studio environment for all the performance evaluation.  For Table~\ref{tab-4} and Fig.~\ref{fig-11} only scenario-2 is employed to observe the robustness of the proposed algorithms. During the LoS data collection, there was absolute clear environment between the anchors and tag. For NLoS data collection, there was an iron sheet placed between the anchors and the tag, so that no direct path of signal can be transmitted or received by it. The tag was connected to a PC and the data was logged via the Teraterm software into a comma-separated value (CSV) file. The collected dataset, had $15000$ signals and $1000$ LoS and $100$ NLoS signals are collected randomly. This selection results in a ratio of $1:0.1$ for the  LoS and NLoS signal. Finally, we randomly select different proportions of LoS and NLoS data to test the robustness of proposed algorithm. 
\begin{figure*}[!ht]
\begin{minipage}[!h]{0.5\linewidth}
\begin{center}
\title*{a. Estimated distance}
\includegraphics[width=1\linewidth]{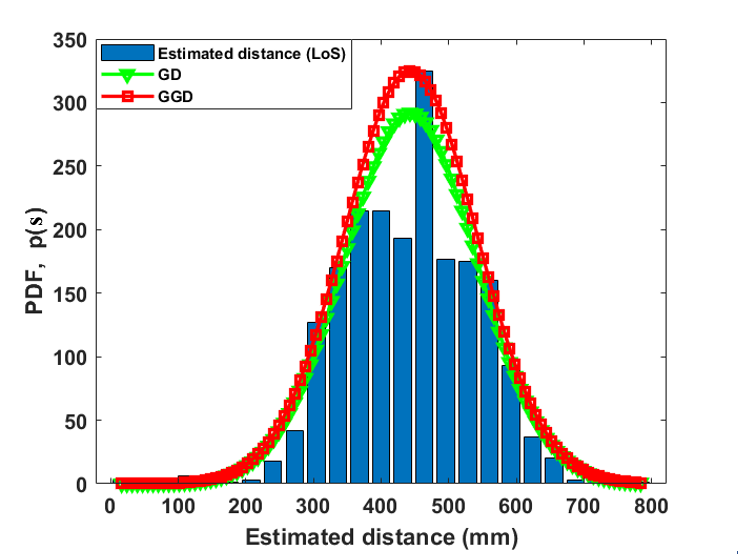}
\end{center}
\end{minipage}
\begin{minipage}[!ht]{0.5\linewidth}
\begin{center}
\title*{b. FP Power Level}
\includegraphics[width=1\linewidth]{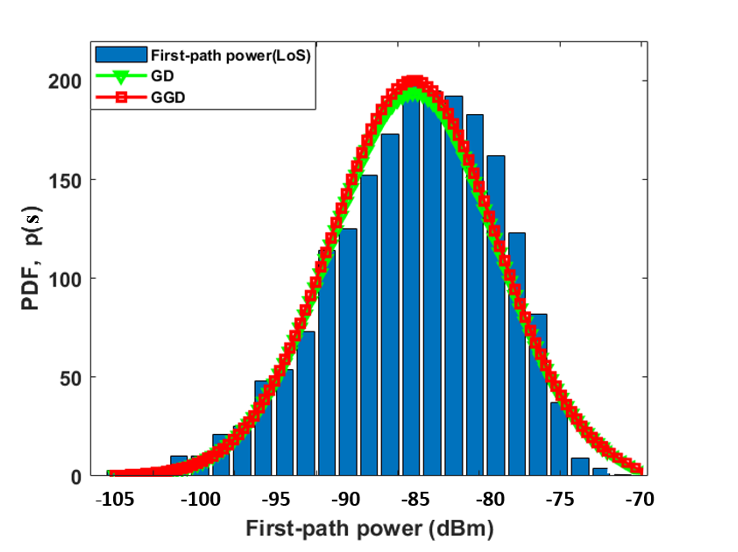}
\end{center}
\end{minipage}
\begin{minipage}[!ht]{0.5\linewidth}
\vspace{1cm}
\begin{center}	
\title*{c. Received Power Level}
\includegraphics[width=1\linewidth]{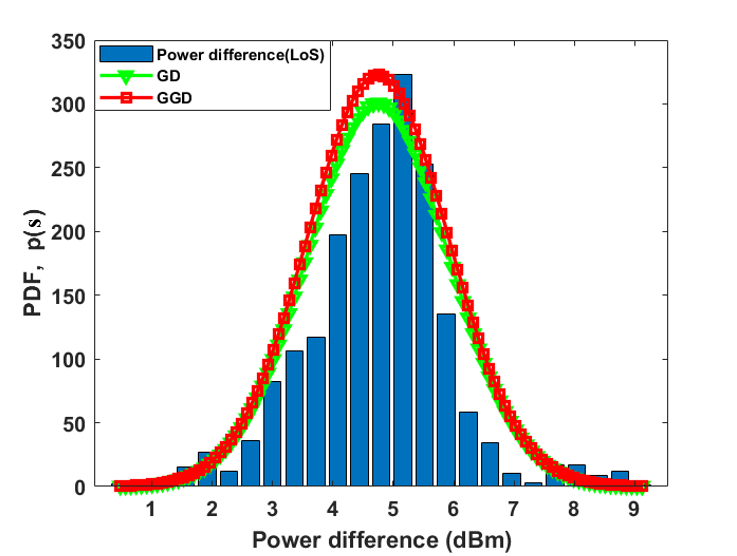}
\end{center}
\end{minipage}
\begin{minipage}[!ht]{0.5\linewidth}
\vspace{1cm}
\begin{center}	
\title*{d. Threshold Power}
\includegraphics[width=1\linewidth]{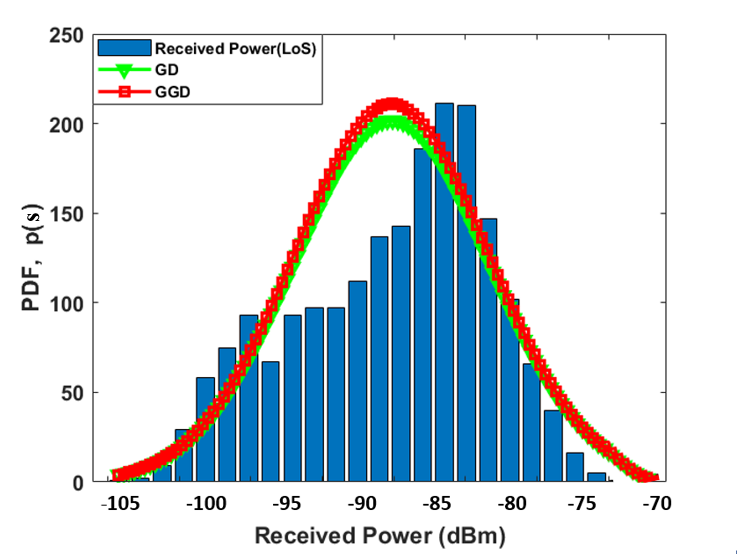}
\end{center}
\end{minipage}
\caption{Histogram distribution of our $4$ key features in LoS environment a) Estimated distance, b) first path power level, c) received power level, and d) the threshold power.}\label{fig-3}
\end{figure*}

\begin{figure*}[!ht]
\begin{minipage}[!h]{0.5\linewidth}
\begin{center}
\title*{a. Estimated distance}
\includegraphics[width=1\linewidth]{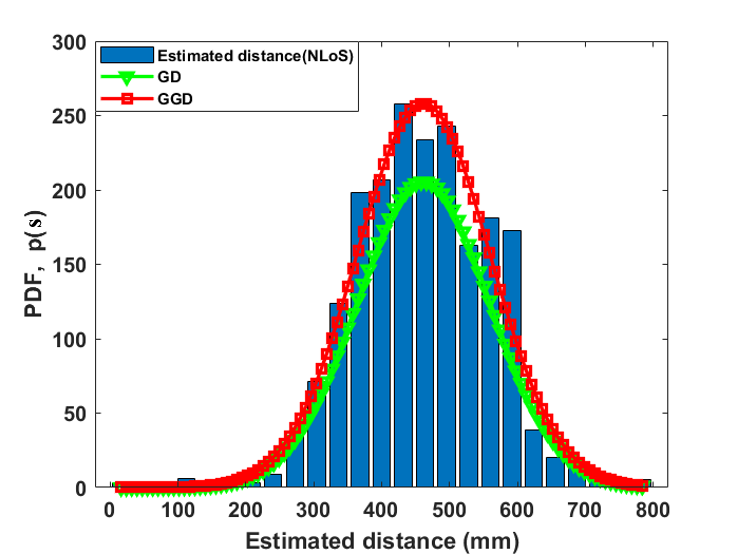}
\end{center}
\end{minipage}
\begin{minipage}[!ht]{0.5\linewidth}
\begin{center}
\title*{b. FP Power Level}
\includegraphics[width=1\linewidth]{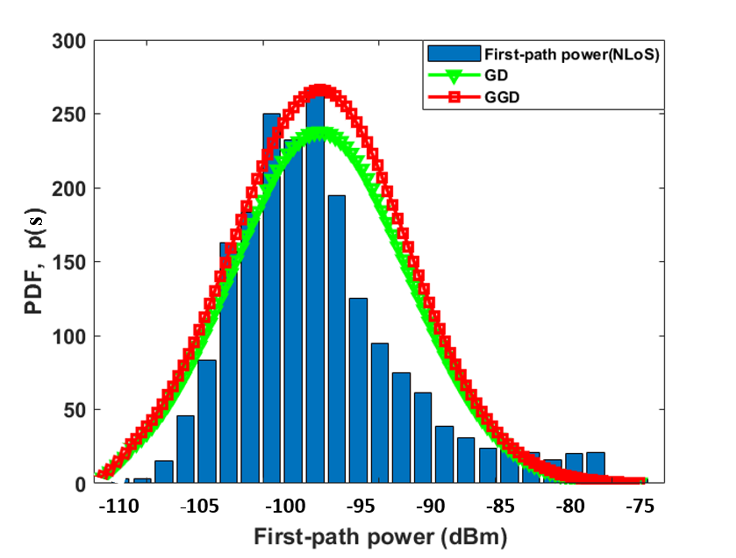}
\end{center}
\end{minipage}
\begin{minipage}[!ht]{0.5\linewidth}
\vspace{1cm}
\begin{center}	
\title*{c. Received Power Level}
\includegraphics[width=1\linewidth]{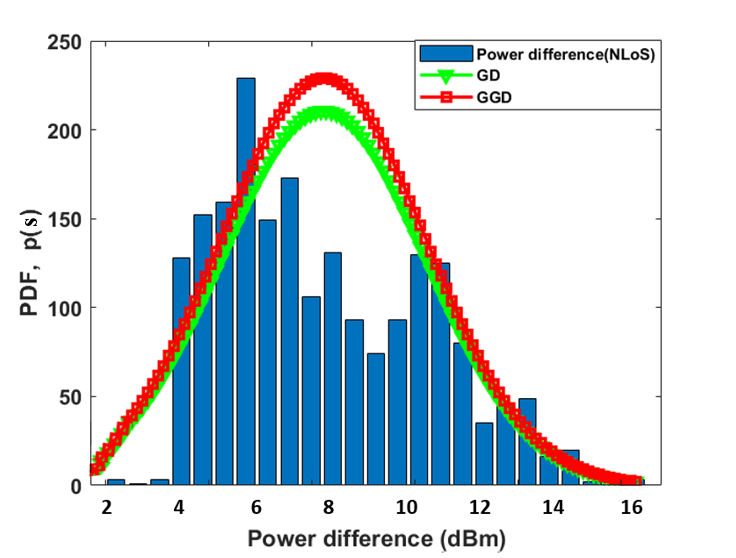}
\end{center}
\end{minipage}
\begin{minipage}[!ht]{0.5\linewidth}
\vspace{1cm}
\begin{center}	
\title*{d. Threshold Power}
\includegraphics[width=1\linewidth]{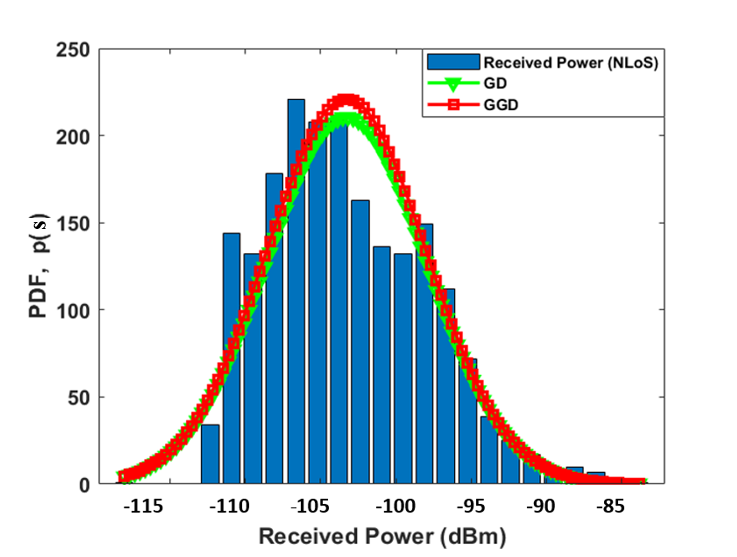}
\end{center}
\end{minipage}
\caption{Histogram distribution of our $4$ key features in NLoS environment a) Estimated distance, b) first path power level, c) received power level, and d) the threshold power.}\label{fig-4}
\end{figure*}
\begin{table}[!t]
\centering
\caption{Configurations of the MDEK-$1001$ UWB kit.}\label{table-1}
\begin{center}
\begin{tabular}{|l|l| }
\hline
\textbf{Properties} & \textbf{Values}  \\
\hline
Chip & DW 1000 \\
\hline
Transceiver& DWM1000 \\
\hline
Pulse shape & Gaussian pulse \\
\hline
Number of Anchors & 4 \\
\hline
Data Rate & 6.8 Mbps \\
\hline
Frequency & 3993.6 MHz \\
\hline
Bandwidth & 499.2 MHz\\
\hline
Channel & 2\\
\hline
Pulse Repetition Frequency (PRF) & 16 MHz\\
\hline
\end{tabular}
\end{center}
\end{table}
For our analysis, $7$ signal components are extracted that are:
\begin{enumerate}
\item Amplitude of the first path ($F_1$).
\item Amplitude of the second path ($F_2$).
\item Amplitude of the third path ($F_3$).
\item Preamble accumulation count value.
\item Amplitude of the channel impulse response (CIR).
\item Standard noise variance reported in the DW-$1000$ chipset.
\item The estimated calculated distance.
\end{enumerate}
In brief, among the mentioned NLoS identification methods in the literature, the threshold difference between the first-path power and received power have been widely used in different ML algorithms~\cite{Sang-2020, Mirama-2021,Che-2020}. In our analysis we will use the above $7$ signal components to calculate our $4$ key features which are the estimated distance (\ref{eq-5}), first path power level~(\ref{eq-17}), received power level~(\ref{eq-18}), and the power difference between the first and the received power level~(\ref{eq-19}). The first-path power level is calculated as~\cite{Che-2020,Decawave}
\begin{equation}\label{eq-17}
\textrm{FP Level}=10 \times \log_{10}\left(\frac{F_1^2+F_2^2+F_3^2}{N^2}\right)-A \quad \mathrm{dBm}, 
\end{equation}
\noindent
where $F_1$, $F_2$ and, $F_3$ represent first, second and third harmonics of the first-path signal amplitudes. $A$ is a constant equivalent to $113.77$ when a PRF is $16$~MHz as mentioned in~\cite{Decawave} (page-46), and $N$ is the Preamble Accumulation Count value. The received power level of the signal can be computed as~\cite{Decawave}
\begin{equation}\label{eq-18}
\textrm{RX Level}=10 \times \log_{10}\left(\frac{CIR \times 2^{17}}{N^2}\right)-A \quad \mathrm{dBm},
\end{equation}
\noindent
where $CIR$ is the Channel Impulse Response Power value \textcolor{blue}{and $2^{17}$ is a correction factor so that the register value of DWM 1000 can be converted into the right magnitude for conversion into dBm.} The power dissipation in NLoS environment is higher than the LoS environment due to multi-path effects, resulting in the first path of the LoS signal to have more power than the the first path of the NLoS signal. This knowledge can now be employed to improve the detection capability of the algorithm, therefore, the difference between the received and first-path power can also be employed. The formula is as shown in 
\begin{equation}\label{eq-19}
\textrm{Threshold Power = RX Level - FP Level}
\end{equation}

Figures~\ref{fig-3} and~\ref{fig-4} show the probability density function (pdf) of the selected four features which are the estimated distance, first-path power, received path power, and the power difference. Histogram function is employed to generate the pdfs of these features and are represented by the blue bars. The GD and GGD distribution is plotted by using the equation~(\ref{eq-9}) and~(\ref{eq-16}), respectively. For GD distribution, only mean and variance of the data is required. However, for GGD distribution, mean, variance, and kurtosis is required. The distribution of the features can be more closely approximated with the GGD as it has three parameters to update as compared to GD which only has two parameters. From these figures, it can be observed that the selected features follow GD and GGD distribution, therefore, the proposed algorithms can be used for the data classification. Let us now look into designing the selection threshold for these pdfs.
\subsection{Threshold Selection, $\epsilon$}
In order to classify the LoS and NLoS signals, we need to select an appropriate threshold $\epsilon$ as mentioned in section~\ref{sec:Section-3}. For threshold $\epsilon$ selection, we start by constructing a training set, then use the remaining LoS and NLoS signals to cross-check and validate the results, and finally carry out the testing as mentioned above. According to the training data set, we estimate the mean, variance, and kurtosis of each features to build function $P(\pmb{s}_i)$ as mentioned in algorithm~\ref{algorithm-1}. The threshold $\epsilon$ was chosen based on the $F$-Score. $F$-Score is defined as the weighted average of precision and recall and calculated as

\begin{figure*}[!ht]
\begin{minipage}[!ht]{0.5\linewidth}
\begin{center}
\title*{a. \textbf{SVM-based algorithm}}
\includegraphics[width=1.0\linewidth]{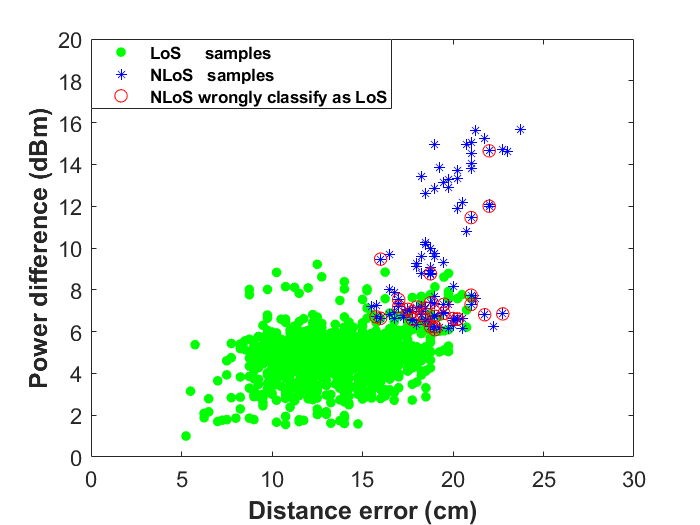}
\end{center}
\end{minipage}
\begin{minipage}[!ht]{0.5\linewidth}
\begin{center}	
\title*{b. \textbf{DT-based algorithm}}
\includegraphics[width=1.0\linewidth]{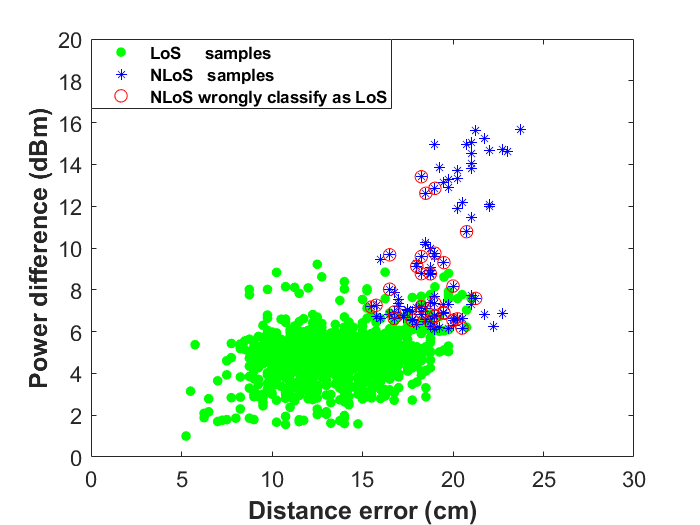}
\end{center}
\end{minipage}
\begin{minipage}[!ht]{0.5\linewidth}
\begin{center}
\title*{c. \textbf{NB-based algorithm}}
\includegraphics[width=1.0\linewidth]{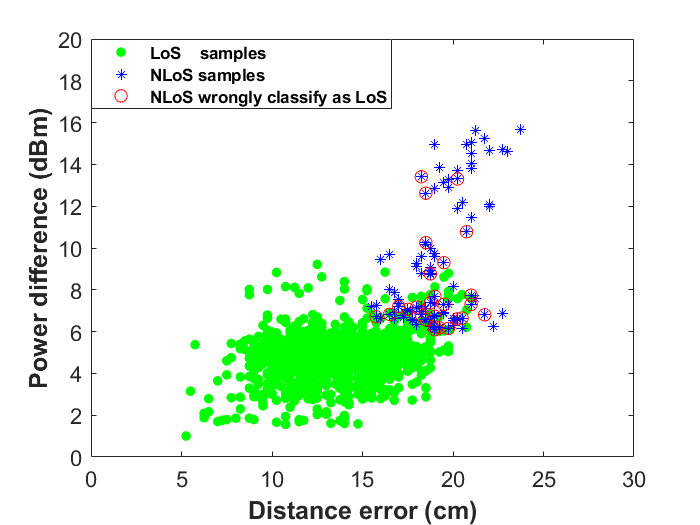}
\end{center}
\end{minipage}
\begin{minipage}[!ht]{0.5\linewidth}
\begin{center}
\title*{d. \textbf{NN-based algorithm}}
\includegraphics[width=1.0\linewidth]{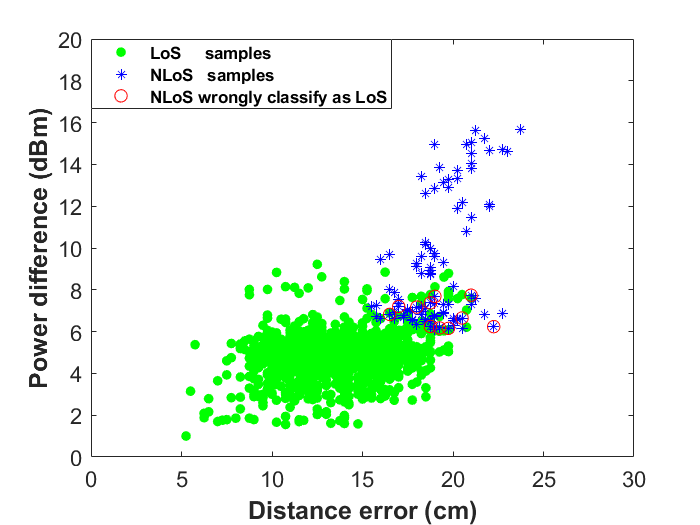}
\end{center}
\end{minipage}
\begin{minipage}[!ht]{0.5\linewidth}
\begin{center}	
\title*{e. \textbf{GD-based algorithm}}
\includegraphics[width=1.0\linewidth]{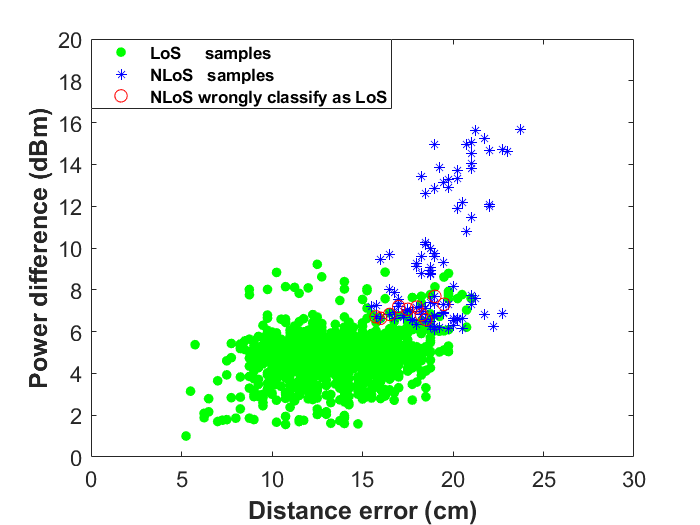}
\end{center}
\end{minipage}
\begin{minipage}[!ht]{0.5\linewidth}
\begin{center}	
\title*{f. \textbf{GGD-based algorithm}}
\includegraphics[width=1.0\linewidth]{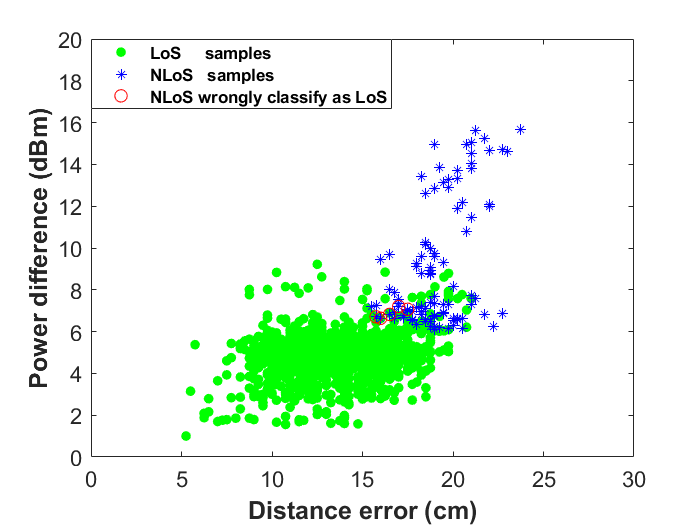}
\end{center}
\end{minipage}
\caption{Visualization of samples a) SVM-, b) DT-, c) NB-, d) NN-, e) GD-, and f) GGD-based algorithms.}~\label{fig-5}
\end{figure*}
\begin{equation}\label{eq-20}
\textrm{F-Score} = \frac {2 \times (\textrm{Recall $\times$ Precision})}{\textrm{Recall $+$ Precision}},
\end{equation}
where Precision and Recall are defined as
\begin{eqnarray}\label{eq-21,22}
\textrm{Precision} &=& \frac{TP}{TP+FP},\\
\textrm{Recall} &=& \frac{TP}{TP + FN},
\end{eqnarray}
where $TP$ is the True Positive, $FP$ is the False Positive, $FN$ is the False Negative, and $TN$ is the True Negative, respectively. $TP$ means that the instances are classified as positive when they are actually positive, $TN$ illustrates the instances are classified as negative when they are in negative condition. $FP$ shows that the instances are classified as positive when they are negative. Similarly, $FN$ represents the instances classified as negative when they are actually positive. However, the indoor environment changes and therefore, we have added a forgetting factor $\lambda$ for the calculation of threshold $\epsilon$. In such a case, the threshold can be updated after training and therefore can incorporate small changes in the environment. The threshold can be updated by the following equation
\begin{equation}\label{eq-23}
\epsilon_{t+1}=\epsilon_t+\lambda \times e_t,
\end{equation}
where $t$ is the time index and $e_t$ is the error if we classify the LoS as NLoS or vice versa. This will help us update the threshold after training.
\begin{figure*}[!ht]
\begin{minipage}[!h]{0.5\linewidth}
\begin{center}
\title*{a. SVM algorithm}
\includegraphics[width=1\linewidth]{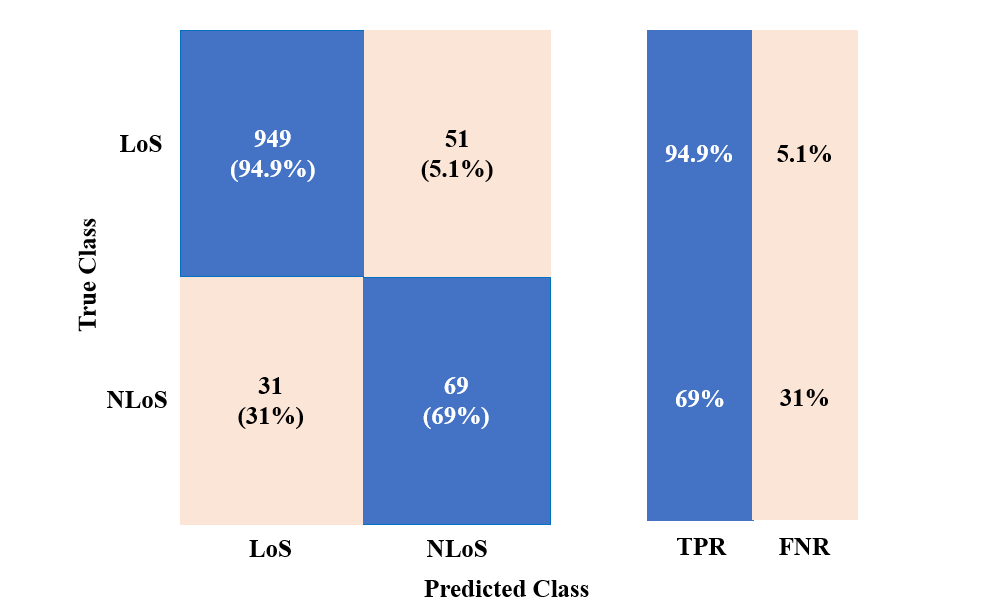}
\end{center}
\end{minipage}
\begin{minipage}[!ht]{0.5\linewidth}
\begin{center}
\title*{b. DT algorithm}
\includegraphics[width=1\linewidth]{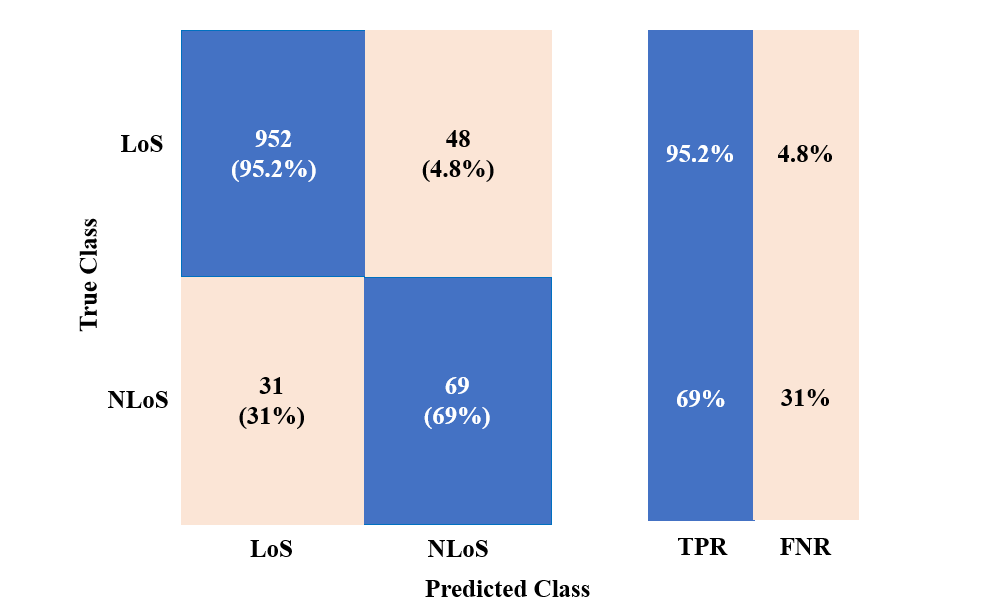}
\end{center}
\end{minipage}
\begin{minipage}[!ht]{0.5\linewidth}
\vspace{1cm}
\begin{center}	
\title*{c. NB algorithm}
\includegraphics[width=1\linewidth]{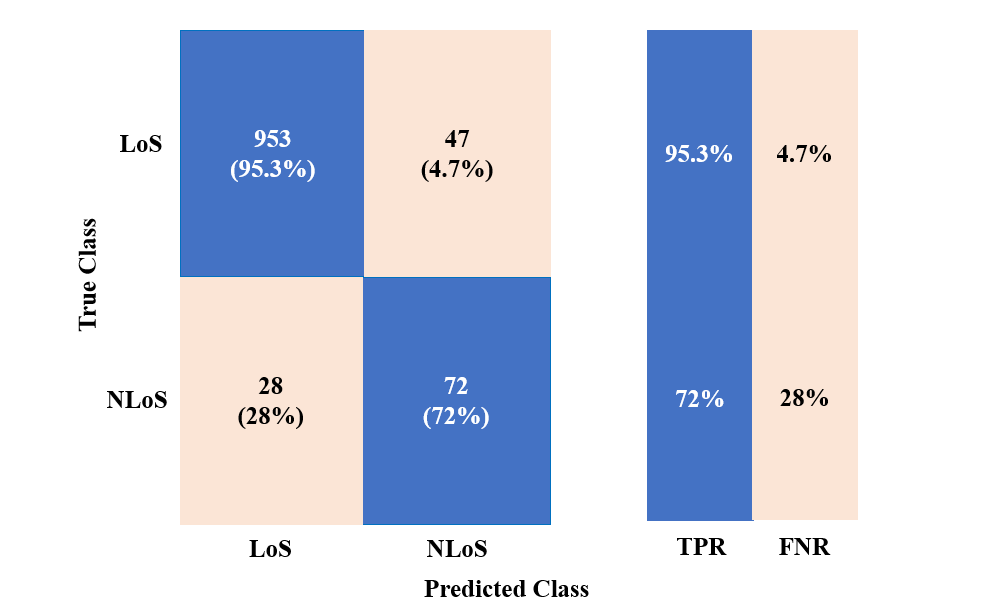}
\end{center}
\end{minipage}
\begin{minipage}[!ht]{0.5\linewidth}
\vspace{1cm}
\begin{center}
\title*{d. NN algorithm}
\includegraphics[width=1\linewidth]{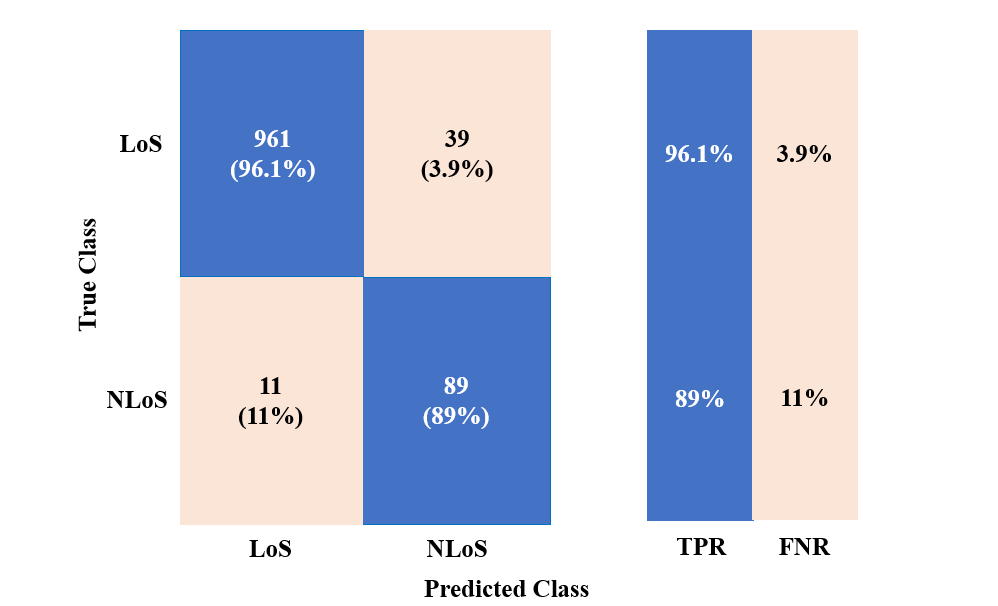}
\end{center}
\end{minipage}
\begin{minipage}[!ht]{0.5\linewidth}
\vspace{1cm}
\begin{center}	
\title*{e. GD algorithm}
\includegraphics[width=1\linewidth]{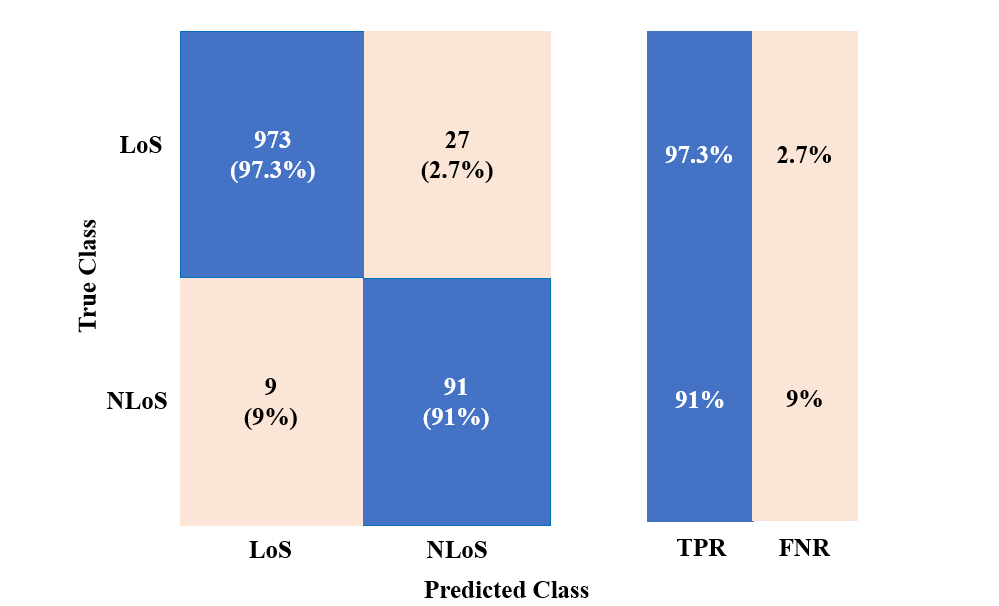}
\end{center}
\end{minipage}
\begin{minipage}[!ht]{0.5\linewidth}
\vspace{1cm}
\begin{center}	
\title*{f. GGD algorithm}
\includegraphics[width=1\linewidth]{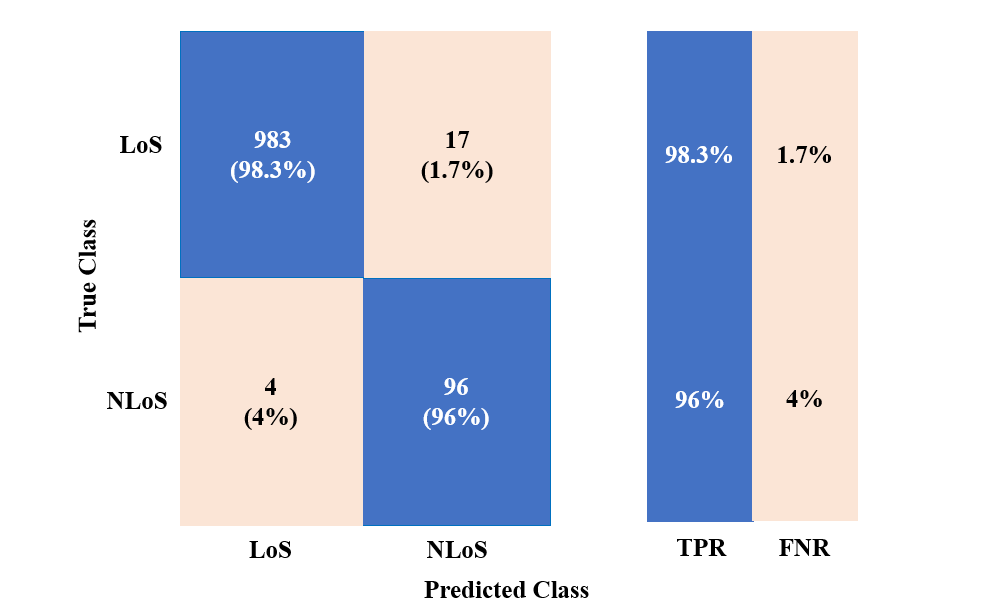}
\end{center}
\end{minipage}
\caption{The confusion matrix of the validation data set on six algorithms a) SVM, b) DT, c) NB, d) NN, e) GD, and f) GGD algorithms. The proposed algorithms improve the classification significantly.}\label{fig-6}
\end{figure*}
\section{Performance Evaluation}~\label{sec:Section-5}
In this section, we examine the performance of proposed algorithms. We first compare the results of the proposed classifier with the state-of-the-art machine learning algorithm based on SVM, DT, NB, and NN and start by computing two quantitative metrics: (i) confusion matrix that shows the classification results of each individual classifier, and (ii) Receiver Operating Characteristics (ROC) curve and corresponding Area-Under-the-ROC curve (AUC) value. Second, we calculate the performance of these algorithms in terms of precision, recall and accuracy. Third, the effect of different ratios of LoS and NLoS is studied for these algorithms and finally, the effect of change in environment to observe the robustness of our proposed algorithms. 

The visualization of the samples are shown in Figure~\ref{fig-5}. The visualization of the samples is plotted with the help of the power difference calculated using~(\ref{eq-19}) and the distance error that is calculated in centimeters. There are $1000$ LoS and $100$ NLoS signals, respectively. In this figure, the green samples represent the LoS signals and the blue colour represents the NLoS signals. The blue samples with red circle are the NLoS samples which have been falsely classified as LoS. It can be observed from these figures that the proposed GGD algorithm can provide a higher classification by setting an appropriate threshold by 
training the features as mentioned in section~\ref{sec:Section-4} as compared to the GD and the classical ML algorithms. For the SVM-, DT-, NB- and NN-based algorithms the performance is relatively poor as compared to GD and the GGD algorithm, this is due to the limited number of NLoS signals in the dataset failed to train a robust model for classification. Finally, it can be concluded that for imbalanced dataset the GD and GGD performs exceptionally better in classifying the LoS and NLoS signals as compared to conventional ML algorithms such as SVM-, DT-, NB-, and NN- based algorithms.
\begin{table}[!h]
\centering
\caption{Dataset of the LoS and NLoS signals}
\begin{tabular}{|l|c|c|c|c|l|}
\hline
\textbf{Signal}&\textbf{Distance}  &\textbf{RX level}  &\textbf{FP level} &\textbf{PD}&\textbf{Classified}\\
&\textbf{m}  &\textbf{dBm}  &\textbf{dBm} &\textbf{dBm}&\\
\hline
LoS &7.92               &-92.72           &-89.63          & 3.09        & LoS  \\
\hline
LoS &9.98               &-94.16            &-89.55           & 4.61     & NLoS \\
\hline
NLoS &7.22               &-95.27            &-88.40           & 6.87     & NLoS  \\
\hline
NLoS & 7.38               &-93.96             &-88.53          & 5.43    &LoS \\
\hline
\end{tabular}
\label{tab-2}
\end{table}

Table~\ref{tab-2} shows an example of LoS and NLoS signal for the proposed UWB system. From the table it can be observed that we have distance and three features that are RX power level, FP power level, and PD power levels followed by how it is classified by the algorithm. From the table and as mentioned previously, the PD is more for NLoS signals as compared to LoS signals. However, still it cannot be simply employed for classification a signal as LoS or NLoS. Therefore, for accurate classification of an UWB system we will require a number of different features.

Figure~\ref{fig-6} plots the confusion matrix of the four known ML algorithms (SVM, DT, ND, and NN) followed by the two proposed algorithms GD and GGD. All these algorithms are based on $1000$ LoS and $100$ NLoS signals. From these confusion metrics, it can be concluded that the worst performance in terms of True Positive Rate (TPR) for LoS components is achieved by SVM which is $94.9\%$. NB algorithm performs $0.01\%$ better than DT in terms of TPR, however, NN algorithm achieves the best performance of $96.1\%$ as compared to traditional ML algorithms. We can observe that the proposed GD and GGD performance is better than the existing ML algorithms and in terms of TPR in LoS components is $97.3\%$ and $98.3\%$, respectively. Similarly, for NLoS components the performance of NN is much superior as compared to the SVM, DT, and NB. SVM and DT can correctly classify only $69\%$ of NLoS components. NB classifies only $72\%$ of TNs. NN can classify $89\%$ while GD and GGD can classify more than $90\%$ NLoS components accurately. Therefore, from Fig.~\ref{fig-6} it can be observed that for both the LoS and NLoS components GD and GGD algorithms perform much superior to the classical ML algorithms such as SVM, DT, NB, and NN.

\begin{figure}[!ht]
    \centering
    \includegraphics[width=1\linewidth]{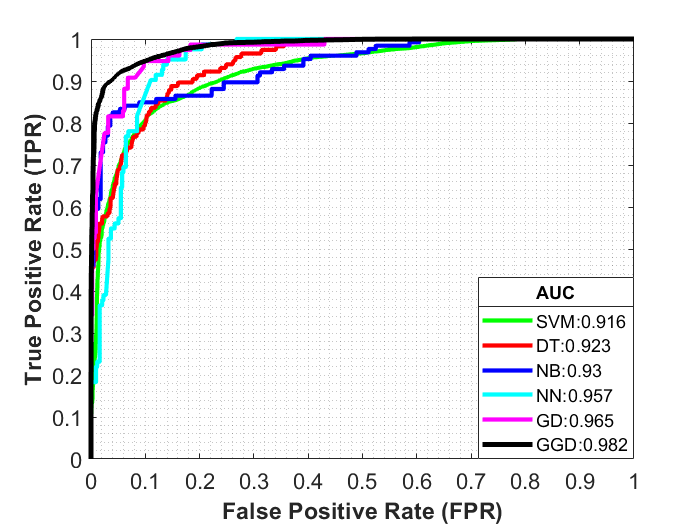}
    \caption{Receiver Operating Characteristics (ROC) and Area under the curve (AUC) comparison of the  algorithms}
    \label{fig-7}
\end{figure}

Figure~\ref{fig-7} plots the receiver operating characteristics (ROC) curve. With these ROC curves, area under the curve (AUC) can be studied for the proposed and ML based positioning algorithms. The ROC curve is plotted with respect to the true positive rate (TPR) versus the false positive rate (FPR). Generally, in a ROC curve the best classifier is closer to the upper left corner, resulting in a larger AUC. From figure~\ref{fig-7} it can be observed that the GGD algorithm is closer to the upper left corner as compared to other algorithms. Furthermore, the AUC of GGD algorithm is $0.982$ which is more than any compared algorithm. As a result, the overall GGD algorithm will be superior in terms of classification accuracy as compared to other algorithms. The second best performance was achieved by the GD algorithm that was around $0.965$ and the SVM performance was the worst despite achieving a value of $0.916$. Out of all the ML algorithms NN performed superior as the area under the curve was equivalent to $0.957$. Finally, it can be observed that the proposed algorithms, therefore GD and GGD, perform superior to the other ML algorithms as they can classify the data more accurately.

\begin{figure}[!ht]
    \centering
    \includegraphics[width=1\linewidth]{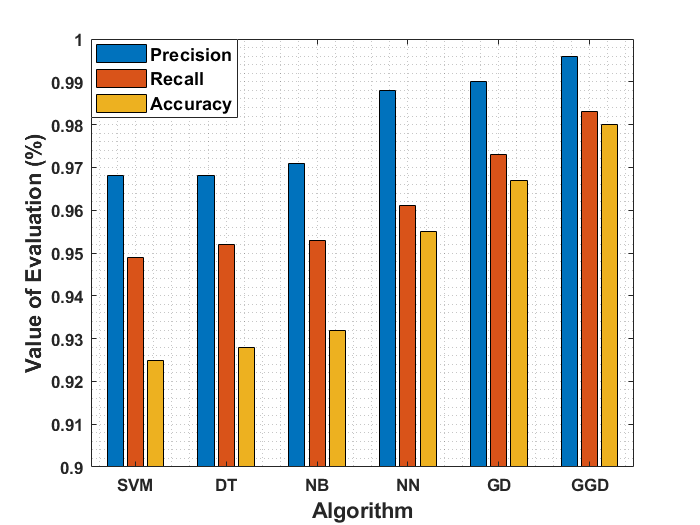}
    \caption{Performance comparison in terms of Precision, Recall, and Accuracy.}
    \label{fig-8}
\end{figure}

Figure~\ref{fig-8} shows the evaluation results of these algorithms in terms of precision, recall, and accuracy.  All these confusion matrices are formed by the help of $1100$ classified samples consisting of $1000$ LoS and $100$ NLoS signals. From the figure, it can be observed that the accuracy of the SVM-based algorithm is equivalent to $92.6\%$. The accuracy of the DT-based, NB-based, and NN-based algorithms is $92.8\%$, $93.2\%$, and $95.5\%$, respectively. The GD-based algorithm results in more than $96.5\%$. The overall accuracy achieved with the GGD-based algorithm is around $98\%$ which shows the GGD-based algorithm is superior to the classical ML and GD-based algorithms. Finally, Table~\ref{tab-3} summarizes and represents the performance of all these algorithms in numbers. It can be observed from the table that GGD algorithm has superior TPR, TNR, Precision, Recall, Accuracy and AUC as compared to all the algorithms. Furthermore, the FPR and FNR are also the lowest as compared to GD and classical ML algorithms. However, in this case, the accuracy of the algorithms is dominated by the LoS signals, therefore, in our next figure we observe the impact of increasing NLoS signals as compared to LoS signals.  
\begin{table*}[!h]
\centering
\caption{Comparison of the proposed GD and GGD with SVM, DT, NB, and NN algorithms.}
\begin{tabular}{|l|c|c|c|c|c|c|c|c|c|}
\hline
\centering
\textbf{Algorithms} &\textbf{TPR} &\textbf{FPR}&\textbf{FNR}&\textbf{TNR}&\textbf{Precision}& \textbf{Recall}& \textbf{Accuracy}& \textbf{AUC}\\
\hline
\textbf{SVM}   & 949   & 51   &31   &69  &0.968  &0.949  &0.926  &0.916\\
\textbf{DT}    & 952   & 48   &31   &69  &0.968  &0.952  &0.928  &0.923\\
\textbf{NB}    & 953   & 47   &28   &72  &0.971  &0.953  &0.932  &0.93\\
\textbf{NN}    & 961   & 39   &11   &89  &0.989  &0.961  &0.955  &0.957\\
\textbf{GD}    & 973   & 27   &9    &91  &0.990  &0.973  &0.967  &0.965\\
\textbf{GGD}   & 983   & 17   &4    &96  &0.995  &0.983  &0.980  &0.982\\
\hline
\end{tabular}
\label{tab-3}
\end{table*}

\begin{figure}[!t]
    \centering
    \includegraphics[width=1\linewidth]{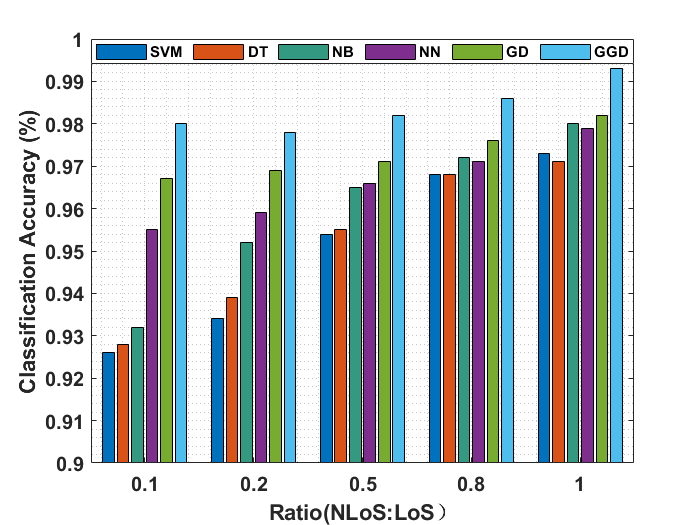}
    \caption{Classification accuracy with different ratios of LoS and NLoS for different algorithms.}
    \label{9}
\end{figure}

Figure~\ref{9} graphically shows the classification accuracy of different ratios of LoS and NLoS samples for the classical ML, GD, and GGD algorithms. The reason for plotting this curve is to see the effect of the imbalanced dataset. From the figure, it can be observed that we have five different ratios $0.1, 0.2, 0.5, 0.8$, and $1.0$, respectively. With a ratio of $0.1$, we will have $1000$ LoS and $100$ NLoS signals. However, as the ratio increases the number of NLoS signals increases. Therefore, for a ratio of $1$, we have $1000$ LoS and $1000$ NLoS signals. From the figure, it can be observed that the classification accuracy significantly improves as the ratio in the imbalanced datasets decreases. For the SVM algorithm, the improvement in classification accuracy performance is significantly more as compared to the other algorithms. For a ratio of $0.1$, the accuracy is around $92.6\%$, while the maximum accuracy is achieved when we have a balanced dataset with a ratio of $1$, therefore, around $97.2\%$. For the GD algorithm, the classification accuracy improves from $96.7\%$ to $98\%$ when the imbalance in the dataset is improved. Finally, for the GGD algorithm, the worst classification is around $98\%$ and the best is around $99.3\%$ for the balanced dataset. The total improvement in classification accuracy is around $3.6\%$ for the SVM algorithm, $2.5\%$ for the GD algorithm, and $1.5\%$ for the GGD algorithm. This indicates that the GGD algorithm is more robust as compared to GD and NB algorithms for a given imbalanced or balanced dataset. Finally, this simulation result proves that the GD and GGD can guarantee better results for NLoS identification under different situations compared to the ML algorithms especially when the dataset is imbalanced.

\begin{figure}[!t]
    \centering
    \includegraphics[width=1.0\linewidth]{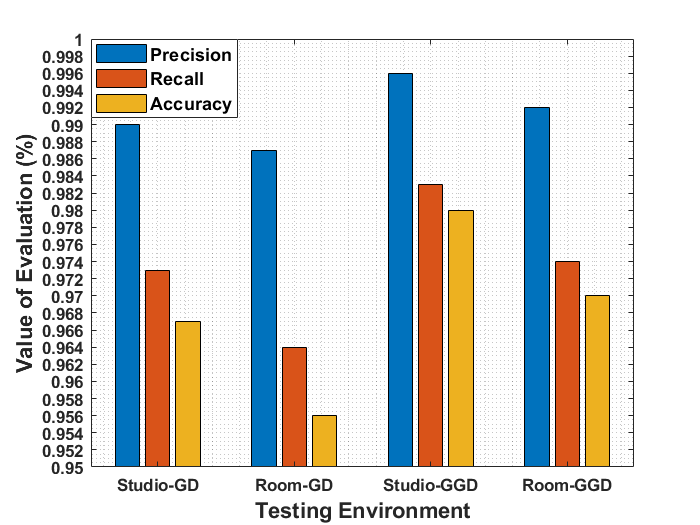}
    \caption{Comparison of static and dynamic threshold $\epsilon$}
    \label{fig-10}
\end{figure}
\begin{table*}[!h]
\centering
\caption{Performance comparison of the two different environments.}
\begin{tabular}{|l|l|l|l|l|l|l|l|l|l|l|}
\hline
\centering
\textbf{Algorithm}&\textbf{Training}&\textbf{Testing} &\textbf{TP} &\textbf{FP}&\textbf{FN}&\textbf{TN}
&\textbf{Precision}&\textbf{Recall}&\textbf{Accuracy}&\textbf{AUC}\\
&\textbf{Scenario}&\textbf{Scenario}&&&&&&&&\\
\hline
\textbf{GD} & Studio   &Studio   & 973   & 27   &9   & 91  & 0.990   & 0.973   &0.967  & 0.965  \\
\hline
\textbf{GD} & Studio   &Room     & 964   & 36   &11  & 89  & 0.987   & 0.964   &0.956  & 0.96 \\
\hline
\textbf{GGD}& Studio   &Studio   & 983   & 17   &4   & 96  & 0.996   & 0.983   &0.98  & 0.982 \\
\hline
\textbf{GGD}& Studio   &Room     & 974   & 26   &7   & 93  & 0.993   & 0.974   &0.97  & 0.975\\
\hline
\end{tabular}
\label{tab-4}
\end{table*}

Figure~\ref{fig-10} shows the impact of the updating the threshold $\epsilon$ after every time instant. There are two approaches shown in the figure that are static and dynamic threshold $\epsilon$ approach.  In the static-threshold approach, the $\epsilon$ is not updated after training and therefore can not update itself if there is any change in the environment. However, in the dynamic threshold approach the $\epsilon$ is updated after training by employing a forgetting factor $\lambda=0.95$ as mentioned in (\ref{eq-23}). It can be observed from the figure that for both the GD and GGD algorithms, the dynamic threshold performs better than the static threshold. An improvement of approximately $0.04$ and $0.02$ is achieved in terms of accuracy by adopting the dynamic threshold approach as compared to static approach for GD and GGD algorithms. Therefore, in this paper dynamic threshold approach is employed.

\begin{figure}[!t]
    \centering
    \includegraphics[width=1\linewidth]{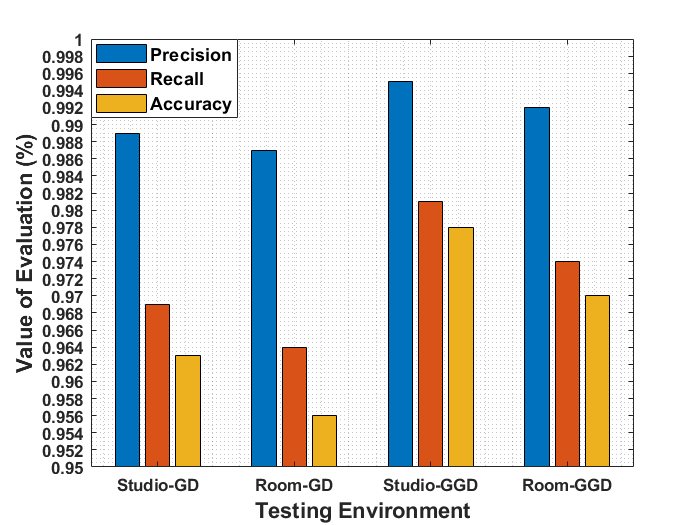}
    \caption{Accuracy results of two different environments}
    \label{fig-11}
\end{figure}
In order to examine the proposed algorithm, we carried out experiments in two different scenarios as shown in Figure~\ref{fig-11}. The configuration of the UWB devices remained the same during the measurements. We build the UWB model using the studio dataset and then tested it on the room data-set. The results are shown in Table~\ref{tab-4} and plotted in Figure~\ref{fig-11}. From Table~\ref{tab-4} and Figure~\ref{fig-11} it can be observed that the performance of GD and GGD-based algorithms is not impacted much by changing the environment. It can be observed that the accuracy of training in the studio and testing at room did not had a big impact. It can be observed from the figure and the table that the accuracy reduced by $0.011$ and $0.01$ when employing the GD- and GGD- based algorithm, respectively. 

\section{Conclusions}~\label{sec:Section-6}
In this paper, a featured-based method for the UWB localisation is proposed. The main aim is to improve the classification of the UWB IPS especially when the dataset is imbalanced, therefore, having a large number of LoS signals as compared to NLoS signals. Initially, in this work seven UWB signal components were collected, and based on these seven signal components, four key features were selected such as estimated distance, first path power level, received power level, and threshold power. With these key features, the joint probability densities for Gaussian and generalised Gaussian are calculated. In order to classify the LoS and NLoS signals a threshold is computed. From the simulation and experimental results, and  it can be observed that the performance of the UWB localisation system was significantly improved by designing the GD and the GGD algorithms as compared to the existing SVM, DT, NB, and NN algorithms. The confusion matrix, ROC, and AUC area for these classification algorithms were compared. In addition, we computed the different ratios of LoS and NLoS signals in the dataset. It can be observed that the classification accuracy improves as the imbalance in dataset is removed, therefore, having same number of LoS and NLoS signals. Finally, from this paper, it can be concluded that the GGD algorithm is effective for NLoS signal identification with balanced Los-NLoS mixed data and remain highly accurate even if we have unbalanced data. For future work, the data can be extended to a large dataset with more signal features to evaluate the classification accuracy. Furthermore, the assumption that all the features are independent will also be removed. 


\balance

\end{document}